\documentclass[conference]{IEEEtran}
\IEEEoverridecommandlockouts

\usepackage{cite}
\usepackage{amsmath,amssymb,amsfonts}
\usepackage{algorithmic}
\usepackage{graphicx}
\usepackage{textcomp}
\usepackage{xcolor}
\usepackage{url}
\usepackage[breaklinks=true]{hyperref}

\usepackage{listings}
\begin{document}

\title{Cumplimiento del Reglamento (UE) 2024/1689 en robótica y sistemas autónomos:\\
una revisión sistemática de la literatura\\
}

\author{\IEEEauthorblockN{1\textsuperscript{st} Yoana Pirta Lorenzo}
\IEEEauthorblockA{\textit{Universidad de León} \\
León, León \\
ypital00@estudiantes.unileon.es}
}

\maketitle

\begin{abstract}
La presente Revisión Sistemática de la Literatura (RSL) tiene por objetivo identificar el estado actual de la adopción y cumplimiento del Reglamento (UE) 2024/1689 en sistemas robóticos y autónomos, analizando herramientas, marcos de ciberseguridad y metodologías aplicadas. Se empleó el protocolo PRISMA, partiendo de 243 registros en IEEE Xplore, ACM DL, Scopus y Web of Science, de los cuales 22 estudios fueron finalmente incluidos tras aplicar criterios de elegibilidad y evaluación de calidad. Los hallazgos revelan que, aunque existen avances en gestión de riesgos y cifrado de comunicaciones, persisten lagunas significativas en módulos de explicabilidad, supervisión humana en tiempo real y trazabilidad de la base de conocimiento. Asimismo, solo el 40\% de las soluciones incorpora de manera explícita los requisitos de transparencia y un 30\% habilita mecanismos de intervención ante fallos. Como conclusión, se constata un cumplimiento parcial de los mandatos del reglamento de IA, subrayando la necesidad de desarrollar enfoques modulares que integren de forma completa los criterios de riesgo, supervisión y auditoría continua en entornos robóticos autónomos.\\[1ex]
\end{abstract}


\section{Introducción}
El Reglamento (UE) 2024/1689 (en lo adelante: reglamento de IA), establece en la Unión Europea un marco normativo esencial para el desarrollo seguro y responsable de tecnologías basadas en inteligencia artificial, clasificando como “alto riesgo” a los sistemas robóticos y autónomos que interactúan directamente con personas y entornos físicos. Sus disposiciones abordan requisitos de gestión de riesgos, transparencia, explicabilidad y supervisión humana continua, con el fin de garantizar la seguridad, la fiabilidad y el respeto a los derechos fundamentales de los ciudadanos.

La convergencia entre robótica autónoma, IA y ciberseguridad ha generado debates y estudios sobre cómo proteger sistemas que aprenden y toman decisiones sin intervención humana directa. Sin embargo, la diversidad de metodologías y la ausencia de criterios estandarizados han provocado que muchas revisiones previas decaigan en cuanto a su aplicabilidad práctica y alineación con las exigencias regulatorias. Estos vacíos dificultan la implementación de soluciones robustas y conformes al reglamento de IA, lo que subraya la necesidad de una revisión sistemática de la literatura que sintetice el estado del arte, identifique tendencias y señale lagunas en la investigación actual sobre el cumplimiento del Reglamento (UE) 2024/1689 en robótica y sistemas autónomos.

\subsection{Justificación de la relevancia del tema elegido}
El avance de los sistemas robóticos y autónomos en los diferentes sectores como la industria manufacturera, la salud y los entornos de interacción humano-robot (HRI) conlleva riesgos considerables en materia de seguridad, privacidad y ética. Estos equipos no solo ejecutan tareas complejas y a menudo peligrosas, sino que también manejan datos sensibles y toman decisiones críticas en tiempo real, lo que los hace un blanco atractivo para ataques cibernéticos y vulnerable a fallos operativos con consecuencias potencialmente graves. En este contexto, el Reglamento (UE) 2024/1689 —el reglamento de IA— impone obligaciones de gestión de riesgos, transparencia y supervisión humana continua, pero su grado de cumplimiento aún no ha sido evaluado de manera exhaustiva.

Aunque existen frameworks consolidados en el ámbito de la robótica, como el Robot Security Framework de Alias Robotics o SROS2 en ROS/ROS 2, estos se centran principalmente en buenas prácticas de auditoría y endurecimiento de la capa de comunicaciones, sin abordar de forma sistemática los requisitos de trazabilidad de decisiones algoritmicas, explicabilidad de los modelos de IA o gobernanza de los flujos de datos impuesta por el reglamento de IA. Esta desconexión entre prácticas de seguridad tradicionales y las nuevas exigencias regulatorias crea vacíos que dificultan la implementación de soluciones que sean, a la vez, robustas y conformes con la normativa europea.

Realizar esta revisión sistemática de la literatura —siguiendo el protocolo PRISMA— aporta la transparencia, la reproducibilidad y el rigor necesarios para sintetizar el estado del arte, evaluar objetivamente hasta qué punto los sistemas robóticos cumplen con el reglamento de IA y detectar las áreas donde se requieren desarrollos adicionales. Los resultados de este análisis no solo orientarán a los investigadores en la identificación de lagunas críticas, sino que también ofrecerán a legisladores y empresas una ruta de prácticas efectivas y recomendaciones para avanzar hacia un ecosistema robótico verdaderamente seguro, responsable y regulado.

\subsection{Definición de la pregunta de investigación}
Para estructurar esta revisión sistemática de acuerdo al Modelo PICO (Población, Intervención, Comparación, Outcome), la investigación se enfocará en robots autónomos dotados de componentes de Inteligencia Artificial (ejemplo, sistemas basados en ROS/ROS 2) como población de estudio. La intervención comprende las herramientas y metodologías de ciberseguridad (frameworks, librerías, plataformas) diseñadas para estos entornos. La comparación se establece frente a la situación en la que dichas soluciones no se encuentran alineadas con los requisitos del Reglamento (UE) 2024/1689. Finalmente, el outcome o resultado esperado es el grado en que esas herramientas cubren los mandatos normativos relacionados con la gestión de riesgos, la auditoría continua, la supervisión humana y la trazabilidad a lo largo de todo el ciclo de vida del robot.
\textbf{Investigación (PICO)}
\begin{itemize}
  \item \textbf{Población}: Robots autónomos con componentes de IA (p. ej., sistemas ROS/ROS 2). 
  \item \textbf{Período}: enero 2018–marzo 2025.  
  \item \textbf{Intervención}: Herramientas y metodologías de ciberseguridad (frameworks, librerías, plataformas).
  \item \textbf{Comparación}: Ausencia de alineación con las obligaciones del Reglamento (UE) 2024/1689.
  \item \textbf{Outcome}: Grado de cobertura de requisitos normativos —gestión de riesgos, auditoría continua, supervisión humana y trazabilidad— a lo largo de todo el ciclo de vida del robot.
\end{itemize}

Durante la investigación se formularon y afinaron un conjunto de preguntas de investigación concretas (PI1–PI4) para acotar el alcance de la revisión sistemática de la literatura y permitir una evaluación rigurosa de los estudios seleccionados. Estas preguntas fueron a base para:
\begin{itemize}
  \item \textbf{PI1.} Identificar y clasificar las principales herramientas, frameworks y bibliotecas de ciberseguridad diseñadas para entornos de robótica autónoma (p. ej., ROS/ROS 2).
\item \textbf{PI2.} Determinar en qué medida dichas soluciones satisfacen los requisitos normativos del Reglamento (UE) 2024/1689 relativos a gestión de riesgos, transparencia de funcionamiento, supervisión humana y trazabilidad.
\item \textbf{PI3.} Detectar lagunas y áreas insuficientemente cubiertas, especialmente en la protección de bases de conocimiento simbólicas y subsimbólicas, auditoría continua y resiliencia frente a ataques adversariales.
\end{itemize}
Con estas preguntas se pretende no solo mapear el estado del arte, sino también señalar líneas de investigación futura orientadas a cerrar los vacíos detectados y facilitar la adopción de buenas prácticas en el desarrollo y despliegue de sistemas robóticos conformes con la Ley de IA europea.

\subsection{Objetivos de la revisión}
El presente estudio tiene como objetivo ofrecer una visión exhaustiva y actualizada del grado de cumplimiento del Reglamento (UE) 2024/1689 en robótica y sistemas autónomos, así como identificar las tendencias y lagunas en las herramientas de ciberseguridad empleadas. Mediante una revisión sistemática de la literatura. 

Con esta base se avaluará en qué medida estas soluciones satisfacen los requisitos de gestión de riesgos, transparencia, supervisión humana y trazabilidad establecidos por el reglamento de IA europeo.

Centrándose principalmente en los siguientes puntos:
\begin{enumerate}
\item Identificar el estado actual del cumplimiento del Reglamento (UE) 2024/1689 en sistemas robóticos y autónomos.

\item Evaluar las metodologías y tecnologías utilizadas en la implementación de los requisitos del reglamento.

\item Analizar las brechas y vacíos en la investigación sobre el cumplimiento de las normativas en la robótica autónoma.

\item Proponer recomendaciones y direcciones futuras para mejorar el cumplimiento y la investigación en este ámbito.
\end{enumerate}

\section{Metodología}
Las referencias bibliográficas consultadas fueron encontradas en las bases de datos IEEE Xplore, ACM Digital Library, Scopus y Web of Science, seleccionadas por su prestigio internacional y por ofrecer contenido especializado en robótica, inteligencia artificial y ciberseguridad. 

Tras definir las plataformas de consulta, se determinaron las palabras clave específicas para enfocar la estrategia de recuperación de información. Los términos empleados fueron: robotics, robot, ROS, autonomous, security, cybersecurity, framework, tool, methodology, EU 2024/1689 y AI. La combinación de estos términos se realizó mediante operadores booleanos (AND y OR), integrando también técnicas de truncamiento y uso de comillas para mejorar la precisión en la recuperación de resultados.
Se siguió la metodología PRISMA \cite{b2,b3}.  
\begin{itemize}
  \item \textbf{Bases de datos}: IEEE Xplore, ACM Digital Library, Scopus, Web of Science.  
  \item \textbf{Período}: enero 2018–marzo 2025.  
  \item \textbf{Palabras clave (título/abstract)}:  
   \begin{lstlisting}
("robotics" OR "robot" OR "ROS" OR "autonomous")
AND ("security" OR "cybersecurity")
AND ("framework" OR "tool" OR "methodology")
AND ("EU 2024/1689" OR "AI Act")
\end{lstlisting}
  \item \textbf{Operadores y filtros}: AND / OR, truncamientos (p.\,ej. auton\*), idioma (inglés, español), tipo de documento (artículos, conferencias, informes técnicos).
\end{itemize}

Adicionalmente, se aplicaron restricciones idiomáticas, limitando la búsqueda a documentos en inglés y español, y se filtraron los tipos de publicación considerados relevantes, como artículos científicos, comunicaciones en congresos y documentos técnicos. El rango temporal definido para la recopilación de estudios comprendió desde 2018 hasta 2025 para procurar que los estudios fueran relevantes y actualizados.

\subsection{Criterios de Inclusión y Exclusión de Estudios}
\textbf{Criterios de Inclusión}:
\begin{itemize}
\item Estudios primarios (artículos, conferencias, informes técnicos, software publicado) con validación empírica. 
\item Descripción de herramientas o frameworks aplicados a robótica/IA autónoma. 
\item Referencia explícita (o extrapolable) a los requisitos de la Ley UE 2024/1689. 
\item Artículos científicos revisados por pares.
\item Estudios que traten el cumplimiento del Reglamento (UE) 2024/1689 en sistemas robóticos o autónomos.
\item Publicaciones que aborden aspectos de ciberseguridad, transparencia, y supervisión humana en el contexto de la robótica autónoma.
\end{itemize}

\textbf{Criterios de Exclusión}:
\begin{itemize}
\item Revisiones sin propuesta técnica concreta. 
\item Enfoques limitados a ciberseguridad general sin componente robótico. 
\item Documentos sin acceso completo o sin datos de evaluación. 
\item Artículos que no se centran en el cumplimiento del reglamento o que no abordan sistemas autónomos o robóticos.
\item Estudios que no hayan sido revisados por pares.
\item Artículos no accesibles en texto completo.
\end{itemize}

De un total de 365 registros iniciales, 22 estudios fueron seleccionados tras aplicar los criterios de inclusión y exclusión.
\begin{figure}[htbp]
\centerline{\includegraphics[width=0.4\textwidth]{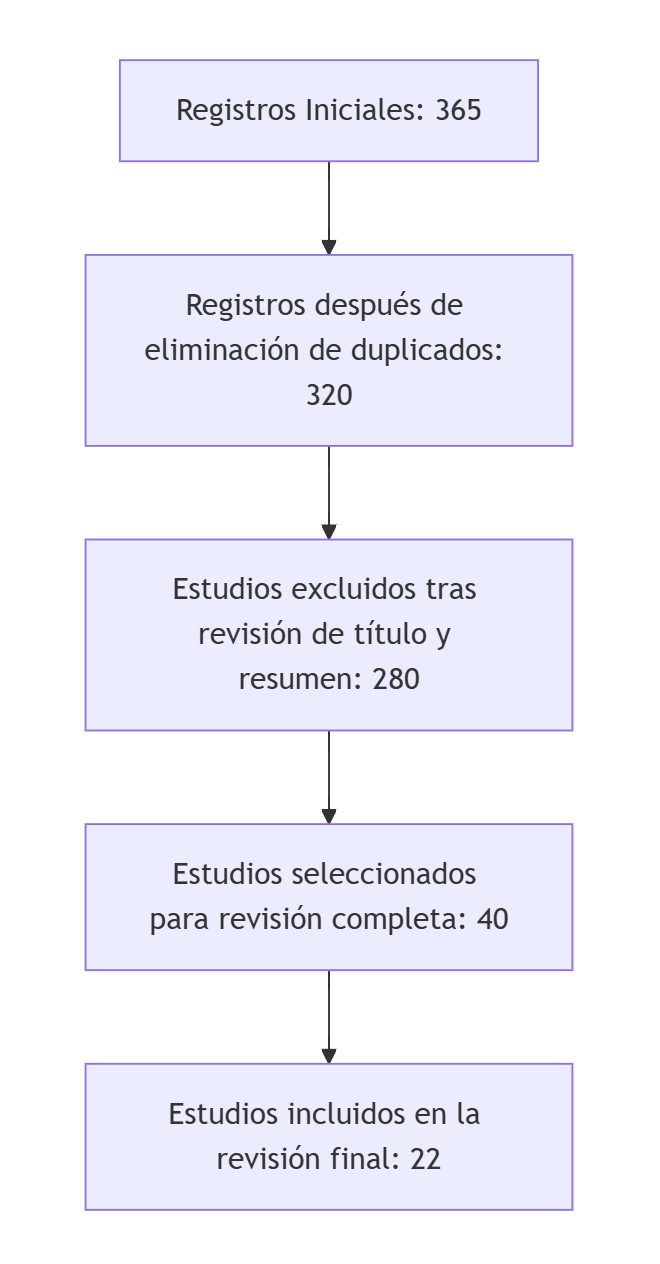}} 
\caption{Diagrama PRISMA}
\label{DiagramaPRISMA}
\end{figure}

\begin{figure}[htbp]
\centerline{\includegraphics{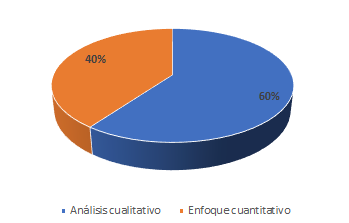}}
\caption{Clasificación de los Estudios Seleccionados Porcentaje}
\label{ClasificacEstudios}
\end{figure}

\subsection{Clasificación por Metodología}
Con base en el análisis de la literatura revisada, se procedió a clasificar los estudios de acuerdo con las metodologías, tecnologías y enfoques que emplearon. Esta categorización proporciona una visión estructurada de las tendencias actuales en el área de ciberseguridad aplicada a sistemas autónomos y robótica, particularmente en relación con el cumplimiento normativo.

Donde, como se muestra en la Figura~\ref{ClasificacEstudios}:
\begin{itemize}
  \item \textbf{El 60\%}: de los estudios seleccionados utilizaron metodologías de tipo cualitativo. Estos trabajos se centraron en evaluaciones sistemáticas, estudios de caso, revisiones normativas y propuestas de marcos conceptuales para medir el cumplimiento de regulaciones como el Reglamento (UE) 2024/1689 o el GDPR en entornos robóticos y de IA.
\item \textbf{El 40\%}:  restante empleó enfoques cuantitativos, que incluyeron simulaciones, métricas de desempeño, validaciones experimentales y estadísticas de cumplimiento de medidas de ciberseguridad. Estos trabajos buscaron medir objetivamente la eficacia de herramientas o marcos de seguridad implementados en entornos autónomos.
\end{itemize}

\subsection{Clasificación por Tecnologías Utilizadas}
\begin{figure}[htbp]
\centerline{\includegraphics[width=0.5\textwidth]{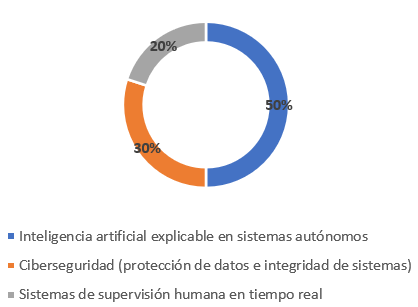}} 
\caption{Clasificación de las tecnologías utilizadas Porcentaje}
\label{TecnologUtilizadas}
\end{figure}

Mientras que al investigar y comparar tecnologías utilizadas se obtuvo que en la Figura~\ref{TecnologUtilizadas} se observa que:
\begin{itemize}
  \item \textbf{El 50\%}: de los estudios abordaron el uso de técnicas de inteligencia artificial explicable para aumentar la confianza y transparencia en los sistemas autónomos, especialmente en decisiones relacionadas con la seguridad y el cumplimiento normativo.
\item \textbf{El 30\%}: de los trabajos se centraron explícitamente en aspectos de ciberseguridad, abordando temas como la protección de datos personales, la detección de intrusiones, y la integridad de las comunicaciones entre robots y sistemas de control.
\item \textbf{El 20\%}:  restante analizó sistemas de supervisión activa de robots autónomos mediante control humano en tiempo real, con énfasis en la intervención oportuna ante riesgos de seguridad o comportamientos anómalos.
\end{itemize}

\subsection{Clasificación por Enfoques Utilizados}
\begin{itemize}
\item \textbf{Basados en marcos normativos}: Muchos de los estudios siguieron un enfoque normativo, alineando sus propuestas o evaluaciones con estándares reconocidos, como el Robot Security Framework (RSF), la ISO/IEC 27001, o las recomendaciones de NIST.

\item \textbf{Enfoques orientados a la implementación práctica}: Otros trabajos adoptaron enfoques prácticos, desarrollando prototipos de sistemas seguros, middleware de comunicaciones cifradas, o arquitecturas de IA explicable para robots autónomos.

\item \textbf{Modelos híbridos}: Finalmente, algunos estudios combinaron enfoques teóricos y prácticos, integrando análisis de riesgos normativos con la validación experimental en plataformas robóticas reales o simuladas (por ejemplo, usando ROS o Fog Robotics).
\end{itemize}

\section{Discusión y conclusiones}
Este estudio ha llevado a cabo una revisión sistemática de la literatura sobre el cumplimiento del Reglamento (UE) 2024/1689 en el ámbito de la robótica y los sistemas autónomos, proporcionando una visión integral de la situación actual, las metodologías empleadas, y los principales desafíos y avances identificados. Los hallazgos más destacados muestran que, a pesar de los progresos en áreas clave como la ciberseguridad y la gestión de riesgos, persisten vacíos importantes en la implementación de ciertos aspectos del reglamento, como la supervisión humana en tiempo real y la trazabilidad de las decisiones de los sistemas autónomos.

\subsection{Comparación con Estudios Previos}
La revisión reveló que, aunque existen avances significativos en áreas como la gestión de riesgos y la ciberseguridad, los estudios actuales aún presentan lagunas en cuanto a la implementación de la supervisión humana en tiempo real y la trazabilidad de la base de conocimiento. Esto coincide con investigaciones previas que subrayan la necesidad de una mayor integración de estos aspectos en los sistemas autónomos. Uno de los principales descubrimientos es que solo una porción de los estudios analizados (alrededor del 40\%) aborda de manera explícita los requisitos de transparencia y explicabilidad en los sistemas robóticos. Esto sugiere que la integración completa de estos principios sigue siendo una tarea pendiente. Asimismo, la falta de mecanismos de intervención ante fallos en muchos de los enfoques revisados resalta la necesidad de fortalecer las estrategias de respuesta ante incidentes.

\textbf{Respondiendo a la primera pregunta de investigación (PI1)}, se pueden identificar tres grandes categorías de soluciones de ciberseguridad en el ecosistema de robótica autónoma (ROS/ROS 2). En primer lugar, los frameworks de evaluación y hardening, como el Robot Security Framework (RSF), ofrecen una metodología estandarizada para auditorías de seguridad en robots, cubriendo desde el análisis de capa física hasta la de aplicación. Asimismo, SROS2, una extensión de ROS 2, aporta mejoras como canales cifrados, autenticación mutua y políticas de control de acceso basadas en certificados, fortaleciendo la seguridad de las comunicaciones. En segundo lugar, las herramientas de monitoreo y detección desempeñan un papel crucial, destacándose los módulos IDS/IPS integrados en ROS, como ROS-IDS, que inspeccionan el tráfico DDS en tiempo real para detectar anomalías de red. Por otro lado, soluciones SIEM basadas en agentes ligeros, como Wazuh, permiten la recopilación y correlación de logs de robots y contenedores para una auditoría continua. Finalmente, las bibliotecas de refuerzo criptográfico y de integridad, como SealFSv2, proporcionan un sistema de ficheros tamper-evident para almacenar configuraciones críticas y modelos, mientras que las firmas digitales de artefactos (herramientas tipo GPG o LUKS) aseguran que los paquetes de modelos o parámetros no hayan sido alterados.

\textbf{En cuanto a la segunda pregunta de investigación (PI2)}, se observa un grado de cobertura variado de los mandatos normativos del Reglamento (UE) 2024/1689. En términos de gestión de riesgos, tanto el RSF como el NIST CSF proporcionan metodologías robustas para la identificación y priorización de riesgos, aunque no incluyen plantillas específicas para arquitecturas cognitivas robóticas. Respecto a la transparencia de funcionamiento, herramientas como SROS2 y las bibliotecas de trazabilidad de ROS 2 (por ejemplo, el registro de nombres de topics y parámetros) facilitan la explicabilidad de los sistemas, pero adolecen de mecanismos más avanzados de “explicación de decisiones IA”, tales como LIME o SHAP, especialmente para comportamientos subsimbólicos. En lo que respecta a la supervisión humana, aunque los frameworks proponen soluciones como “kill-switch” y “heartbeat” supervisados por operadores, aún carecen de interfaces de control en tiempo real que se integren adecuadamente con sistemas de interacción humano-robot (HRI), lo cual es esencial para intervenir directamente en la capa deliberativa de los sistemas autónomos. Finalmente, en términos de trazabilidad, las soluciones SIEM y SealFSv2 cubren aspectos como el registro de eventos y el versionado de modelos, aunque no existen librerías específicas de ROS que unan de manera automática las trazas de sensores, decisiones de IA y logs de seguridad.

\textbf{En relación con la tercera pregunta de investigación (PI3)}, se identifican varios vacíos y áreas insuficientemente cubiertas. Primero, en cuanto a la protección de bases de conocimiento simbólicas (como PDDL o ontologías), no existen bibliotecas estándar que cifren o firmen dinámicamente los ficheros de dominio y problema, lo que deja expuestos los \textquotedblleft planes\textquotedblright\ del robot a manipulaciones. En cuanto a las bases subsimbólicas (modelos de machine learning y LLMs), falta la implementación de pipelines automáticos para pruebas adversariales en tiempo de despliegue, así como mecanismos de \textquotedblleft fallback\textquotedblright\ seguros en caso de drifts de datos o ataques de envenenamiento. En el ámbito de la auditoría continua, aunque los SIEM cubren los logs de sistema, sigue existiendo un vacío en herramientas que correlacionen los eventos de seguridad con las métricas de desempeño robótico, como los fallos de misión vinculados con incidentes de red. Por último, en lo que respecta a la resiliencia frente a ataques adversariales, faltan librerías en ROS que permitan simular y endurecer modelos de visión, como YOLO\_ROS, o de voz, como Whisper\_ROS, frente a ataques de input crafting, y no se han integrado técnicas de robust training en los frameworks de despliegue de robots.

En conjunto, estos hallazgos subrayan la necesidad urgente de avanzar en soluciones específicas que no solo protejan la infraestructura tradicional de ROS/ROS 2, sino que también aborden los desafíos únicos de las arquitecturas cognitivas y de IA desplegada en cumplimiento con el AI Act.

\subsection{Fortalezas y Limitaciones de los Estudios Analizados}
Entre las principales fortalezas, destaca que la mayoría de los estudios abordan cuestiones clave del cumplimiento normativo, como la transparencia, la explicabilidad de la IA y la ciberseguridad. Además, se emplean enfoques robustos para evaluar las tecnologías y su alineación con los requisitos del AI Act. Sin embargo, también se identificaron importantes limitaciones: la mayoría de las investigaciones se enfocan en soluciones parciales y no ofrecen una visión integral del cumplimiento normativo. La falta de consenso en los criterios de evaluación y la diversidad de enfoques metodológicos dificultan, además, la comparación directa de los resultados.

\subsection{Direcciones para Futuras Investigaciones}
La presente revisión contribuye a una mayor comprensión de cómo los sistemas robóticos y autónomos están alineándose con las exigencias regulatorias, pero también subraya la necesidad de continuar investigando las áreas menos desarrolladas, como la integración modular de los principios regulatorios y la mejora de las metodologías de evaluación del cumplimiento normativo. Es necesario desarrollar enfoques más modulares que integren todos los aspectos del Reglamento (UE) 2024/1689 en los sistemas robóticos y autónomos. Se recomienda, además, realizar más investigaciones sobre la trazabilidad y la auditoría continua de los sistemas autónomos para garantizar un cumplimiento constante de los requisitos legales. Finalmente, futuras investigaciones deberían enfocarse en desarrollar soluciones tecnológicas y marcos regulatorios más completos, que no solo garanticen el cumplimiento, sino que también aborden de manera integral los aspectos éticos y de responsabilidad asociados con el uso de sistemas autónomos en entornos críticos.

\section*{Conclusiones}

El análisis de la literatura muestra que, aunque existen avances significativos en la implementación de ciertos requisitos del Reglamento (UE) 2024/1689, especialmente en áreas como la gestión de riesgos y la ciberseguridad, persisten importantes vacíos en otros aspectos críticos. En particular, la supervisión humana en tiempo real, la transparencia y la trazabilidad siguen siendo retos pendientes de superar. De hecho, solo el 40\% de las soluciones analizadas incluyen explícitamente mecanismos que satisfacen los requisitos de transparencia establecidos en el AI Act, lo que revela una brecha considerable en la adecuación normativa de muchos sistemas autónomos y robóticos. Además, aunque se han registrado avances en ciberseguridad, la literatura señala la necesidad urgente de integrar de manera modular los criterios del AI Act en las arquitecturas de estos sistemas, permitiendo una adaptación más flexible y eficaz. En consecuencia, es fundamental que futuras investigaciones se enfoquen en estas áreas prioritarias, no solo para fortalecer el cumplimiento regulatorio, sino también para impulsar una implementación práctica y segura de la inteligencia artificial en entornos dinámicos y críticos.

\section*{Agradecimientos}
Gracias al Proyecto DMARCE - PID2021-126592OB-C21, PID2021-126592OB-C22 financiado por MICIU/AEI/10.13039/501100011033 y por FEDER, UE.

\end{document}